\documentclass[prb,showpacs,nofootinbib,twocolumn]{revtex4}

\begin{document}

\title{Is the Quantum Hall Effect influenced by the gravitational
  field?}

\author{Friedrich W.\ Hehl$^{1,2}$\email{hehl@thp.uni-koeln.de}, Yuri
  N.\ Obukhov$^{1,3}$\email{yo@thp.uni-koeln.de}, and Bernd
  Rosenow$^1$\email{rosenow@thp.uni-koeln.de}}

\address{$^1$ Institut f\"ur Theoretische Physik, Universit\"at zu
  K\"oln, 50923 K\"oln, Germany   \\
  $^2$ Department of Physics and Astronomy, University of
  Missouri-Columbia, Columbia, MO 65211, USA\\
$^3$ Department of Theoretical Physics,
  Moscow State University, 117234 Moscow, Russia}

\date{13 October 2003}

\begin{abstract}
  Most of the experiments on the quantum Hall effect (QHE) were made
  at approximately the same height above sea level. A future
  international comparison will determine whether the gravitational
  field $\mathbf{g}(x)$ influences the QHE. In the realm of (1 +
  2)-dimensional phenomenological macroscopic electrodynamics, the
  Ohm-Hall law is metric independent (`topological'). This suggests
  that it does not couple to $\mathbf{g}(x)$.  We corroborate this
  result by a microscopic calculation of the Hall conductance in the
  presence of a post-Newtonian gravitational field.
\end{abstract}


%\keywords{Quantum Hall effect, gravitation, premetric electrodynamics}

\maketitle

An experiment done in Grenoble, France, on the Quantum Hall Effect
(QHE) --- does it yield the same result as a corresponding experiment
done in Boulder, Colorado? Recall that in Grenoble the height above
sea level is about 220 m whereas Boulder lies at about 1600 m.  For
two atomic clocks situated in Grenoble and Boulder, respectively, this
difference in heights, and thus the difference in the gravitational
potential, yields a measurable effect on the time keeping process
\cite{Will}. The clock in Grenoble runs slow as compared to the Boulder 
clock.  Accordingly, the influence of the {\it gravitational} field on
effects in {\it atomic} physics is an established fact and it seems
legitimate to ask whether the QHE is also affected by the
gravitational field $\mathbf{g}=\nabla \varphi$, with $\varphi$ as
potential. The Hall conductance may then depend on the dimensionless
quantity $\varphi/c^2$.

The QHE is a fascinating manifestation of quantum mechanics on the
macroscopic level\cite{prangegirvin,Hajdu}. An important ingredient of
the theoretical explanation of the QHE is the idea that the quantized
Hall conductance can be linked to a topological invariant, the Chern
number\cite{Thouless,physicstoday}. The topological interpretation suggests
that the Quantum Hall Resistance (QHR) should be very robust against
perturbations. Indeed, its excellent reproducibility makes the QHR
very suitable for metrological
purposes.\cite{FlowersPetley,MohrTaylor,Uzan,Honnef} As pointed out by
Jeckelmann and Jeanneret\cite{JeckelmannJeanneret}, the
reproducibility of the QHR has been established at different locations
with a precision of about $5\times 10^{-8}$. However, the locations at
which these QHRs were measured are all at about the same height,
namely Gaithersburg, USA (NIST), London (NPL), Sydney
(NML)\cite{Jeckelmann}.  Therefore, in future one should compare the
results on the QHR as a function of height. This is what we would like
to suggest.

What outcome do we expect? We will advance a macroscopic argument that
the QHE and the QHR should be completely independent of the
gravitational potential and will present a microscopic calculation in
support of this view.

{\bf Conductor in a gravitational field.} The idea that the
gravitational field may affect the conductive properties of matter, in
particular those of normal metals or superconductors, is a direct
consequence of the fact that charges (here electrons and ions) carry
mass and energy. Since gravity is universally coupled to the
energy-momentum of matter, it also acts on the electric currents and
the electromagnetic fields in conductors. Because of the equivalence
principle, the same qualitative effects should be caused by
gravitational as well as by inertial forces. This was analyzed for
accelerating (rotating) conductors,\cite{noninert,super1,Ohmlaw} and
for the gravitational analog of the Hall effect.\cite{gravhall}

Electrodynamics in a four-dimensional spacetime ``feels'' inertial and
gravitational effects via the metric-dependent constitutive (spacetime
and material) relations. For conductors, this is manifest in the
covariant generalization of Ohm's law \cite{Ohmlaw}. We can use a
macroscopic phenomenological picture in order to estimate the possible
magnitude of the gravitational effects. For isotropic matter with
conductance $\sigma$, which is at rest in a Cartesian reference frame,
the electric current density 3-vector $j^a$ of the free charges, with
$a=1,2,3$, is related to the electric field $E_a$ by means of Ohm's
law
%
%*************************************************************************
\begin{equation}
  j^a = \sigma\,g^{ab}E_a\,.\label{Ohm}
\end{equation}
%************************************************************************
%
Here $g^{ab}$ is the spatial part of the 4-dimensional spacetime
metric $g^{ij}$. In the gravitational Schwarzschild field of the Earth
with mass $M$, we have
%
%*********************************************************************
\begin{equation}
  g^{ab}=\left(1 + {\frac {GM}{2c^2r}}\right)^{-4}\delta^{ab} \approx
  (1 - 2\varphi/c^2)\,\delta^{ab}\,,
\end{equation}
%**********************************************************************
%
with $\varphi = GM/r$. Consequently, when the electric field and the
current are measured for a conductor not in a flat space with the
Euclidean 3-metric $g^{ab} = \delta^{ab}$, but in a curved spacetime
of the Earth, the classical longitudinal conductance will be modified
by the gravitational field to the effective conductivity $\sigma' = (1
- 2\varphi/c^2)\,\sigma$.  The resulting effect with a relative change
of about $10^{-9}$ is close to the present accuracy of quantum Hall
measurements.

This estimate is based on the macroscopic approach. Another possible
manifestation of gravity in conductors arises from the microscopic
analysis of the redistribution of charges in conducting matter under
the influence of the gravitational or inertial forces. Such a
redistribution leads to the emergence of weak electric and magnetic
fields near the surface of and inside metallic
bodies\cite{induced,review}. For a conductor in a gravitational field
$\mathbf{g}$, the resulting induced electric field is $\mathbf{E}\sim
- 0.1\,(m_i/e)\mathbf{g}$, where $e$ is the elementary charge and
$m_i$ the mass of the ion in the conductor's lattice.

%%%%%%%%%%%%%%%%%%%%%%%%%%%%%%%%%%%%%%%%%%%%%
{\bf Electrodynamics in 1 + 2 dimensions.}
%%%%%%%%%%%%%%%%%%%%%%%%%%%%%%%%%%%%%%%%%%%%%
Since the 1960s, experimentalists were able to create a {\em
2-dimensional electron gas} (2DEG) in suitable transistors at
sufficiently low temperatures and to position it in a strong external
transversal magnetic field $B$.  Then, the electrons can only move in
a plane transverse to $B$ and one space dimension can be suppressed.
Therefore a $(1+2)$-dimensional version of Maxwell's equations would
be appropriate for describing this situation.

We start from the Maxwell equations valid in any spacetime dimension.
In exterior calculus, they can be given compactly as\cite{birkbook}
$dG=J$ and $dF=0$, with $G$ as electromagnetic excitation, $J$ as
electric current, and $F$ as electromagnetic field strength. In tensor
language,\cite{Schouten} they read
%
%**************************  nMax    *******************************
\begin{equation}\label{nMax}
\partial_k {\cal G}^{ik}=  {\cal J}^i\,, \qquad
  \partial_{i}F_{k\ell}+\partial_{k}F_{\ell
i}+\partial_{\ell}F_{ki}=0\,,
\end{equation}
%**********************************************************************
%
with $ {\cal G}^{ik}=- {\cal G}^{ki}$ and $F_{ik}=-F_{ki}$. This
``premetric'' form of the Maxwell equations is totally independent of
the metric.  As a consequence, all field quantities have different
transformation properties: The ${\cal J}^i$ is a contravariant
3-vector density, ${\cal G}^{ik}$ an antisymmetric contravariant
3-tensor density, and $F_{ik}$ a antisymmetric covariant 3-tensor.

In $1+2$ spacetime dimensions, the indices $i,k,\ell$ in (\ref{nMax})
run from $0$ to $2$. Then the currents and the fields can be expressed
in terms of their 2-dimensional quantities,
%
%************************** matrices *********************************
\begin{eqnarray}\label{matrices}
({\cal J}^i)=%\pmatrix{{\cal J}^0\cr {\cal J}^1\cr {\cal J}^2} =
\pmatrix{\rho\cr j^1\cr j^2} \,,\quad
({\cal G}^{ik})&=&%\pmatrix{0&{\cal G}^{01}&{\cal G}^{02}\cr
%-{\cal G}^{01}& 0&{\cal G}^{12}\cr
%-{\cal G}^{02}&-{\cal G}^{12}&0}=
\pmatrix{0&{\cal D}^{1}&{\cal D}^{2}\cr
-{\cal D}^{1}& 0&-{\cal H}\cr
-{\cal D}^{2}&{\cal H}&0}\,,\quad\nonumber \\
  (F_{ik})&=&%\pmatrix{0&F_{01}&F_{02}\cr -F_{01}& 0&F_{12}\cr
%    -F_{02}&-F_{12}&0}=
  \pmatrix{0&-E_1&-E_{2}\cr E_{1}& 0&B\cr E_{2}&-B&0}\,,
\end{eqnarray}
%*********************************************************************
%
with the area densities of charge $\rho$ and current $(j^1,j^2)$, the
magnetic and electric excitations ${\cal H}$ and $({\cal D}^1,{\cal
D}^2)$, and electric and magnetic field strengths $(E_1,E_2)$ and $B$,
respectively.

We substitute (\ref{matrices}) into (\ref{nMax}) and find
%
%*******************************  3max1  *******************************
\begin{eqnarray}\label{3max1}
  \partial_1{\cal D}^1+\partial_2 {\cal D}^2 = \rho\,,\hspace{-10pt}&&\;
 - \partial_2{\cal H}- \dot{{\cal D}}^{1} = j^1\,,\;
  \partial_1{\cal H}- \dot{{\cal D}}^2 = j^2\,,\nonumber\\
\label{3max4}
&&  \partial_1E_2-\partial_2E_1 + \dot{B} = 0\,.
\end{eqnarray}
%***********************************************************************
%
The ${\rm div} B=0$ equation is degenerate and drops out. We assumed
an infinite extension of the 2DEG. If this is no longer a valid
approximation, one has to allow for line currents at the boundary of
the sample (``{edge} currents'') in order to fulfill the Maxwell
equations.

Being interested in the phenomenology of the QHE, we have to connect
in (\ref{nMax}) somehow the current ${\cal J}^i$ with the field
strength $F_{ik}$ by a constitutive law. The linear ansatz of the
Ohm-Hall law,\cite{Frohlich,birkbook}
%
%***************************** xym ************************************
\begin{equation}\label{xym}
{\cal J}^i=\sigma^{ik\ell}F_{k\ell}
\end{equation}
%***********************************************************************
%
links both quantities in a generally covariant form, provided the Hall
conductance $\sigma^{ik\ell}=-\sigma^{i\ell k}$ is a contravariant
3-tensor density of rank 3. The totally antisymmetric Levi Civita
symbol $\epsilon^{ik\ell}=\pm 1,0$ has the same transformation
property as a tensor density.

Accordingly, if we assume isotropy in 3 dimensions, we have, with the
scalar field $\sigma_{\rm H}$,
%
%************************** OhmHalliso *********************************
\begin{equation}\label{OhmHalliso}
{\cal J}^i=%\frac 12
\sigma_{\rm H}\,\epsilon^{ik\ell}F_{k\ell}/2
\end{equation}
%***********************************************************************
%
or, decomposed in time and space,
%
%**********************************************************************
\begin{equation}
\rho=\sigma_{\rm H}\,B\,,\quad
j^1=\sigma_{\rm H}\,E_{2}\,,\quad
j^2=-\sigma_{\rm H}\,E_{1}\,.
\end{equation}
%********************************************************************

These are classical phenomenological laws which need some
interpretation when applied to the description of a quantum system.
Within a classical electron model for the conductivity, one finds for
the Hall conductivity $\sigma_{\rm H}= \frac{\rho}{B}$. This relation
is expressed in (\ref{OhmHalliso}) in generally covariant form with a
scalar conductivity $\sigma_{\rm H}$ and with the additional
information of a vanishing longitudinal conductivity. However, a
vanishing longitudinal conductivity is found in the plateau region of
a quantum Hall system, where the Hall conductivity is given by the
classical value for a charge density $\rho = {1 \over N}{e^2 \over h}
B$, which corresponds to $N$ completely filled Landau levels.  The
quantum mechanical input is the robustness of this phenomenology
against the influence of disorder and against density variations.
With this additional input, the complete independence of the quantum
Hall resistance of the gravitational field can be concluded from
(\ref{OhmHalliso}).

We differentiate (\ref{OhmHalliso}) by $\partial_i$. Then, because of
$\partial_i{\cal J}^i=0$ and $\epsilon^{ik\ell}\partial_iF_{k\ell}=0$,
%
%************************** constant ********************************
\vspace{-5pt}\begin{equation}\label{constant}
  \partial_i\sigma_{\rm H}=0\,,
\end{equation}
%********************************************************************
%
that is, the Hall conductance is constant in time and space. In turn,
if we substitute (\ref{OhmHalliso}) into the right-hand-side of
$\partial_k {\cal G}^{ik}= {\cal J}^i$, it can be integrated and yields
%
%***************************** constant 1 **************************
\begin{equation}\label{constant1}
{\cal G}^{ik}=\epsilon^{ik\ell}\sigma_{\rm H}A_\ell\,,
\end{equation}
%****************************************************************
%
with the electromagnetic potential defined by
$F_{k\ell}=\partial_kA_\ell- \partial_\ell A_k$. The integration
constant is irrelevant since the potential is not uniquely determined
anyway.  Eq.(\ref{xym}) is a remarkable constitutive law that is in
clear contrast to the standard ${\cal
G}^{ik}=\sqrt{-g}g^{i\ell}g^{km}F_{\ell m}$ law of (1+3)-dimensional
vacuum electrodynamics. Eq.(\ref{constant1}) represents a
3-dimensional Chern-Simons electrodynamics.
\medskip

{\bf Microscopic analysis of the gravitational field dependence of the
  QHE.} The theoretical analysis presented in this section is
concerned with the integer QHE only, whereas the macroscopic argument
discussed in the previous section is applicable to both, the integer
and fractional QHE.  A magnetic field perpendicular to a
two-dimensional electron system leads to a quantization of states in
Landau levels (LLs) at energies $E_n = (n + 1/2) \hbar \omega_c$, with
$\omega_c = e B/m$. The density of states (DoS) is $\rho(E) = \sum_n
\delta (E - E_n) \ B/ \Phi_0 $, where the last factor describes the
macroscopic degeneracy of an LL.  Here, $\Phi_0 = h/e$ denotes the
magnetic flux quantum.  Thus, a system with a completely filled
highest LL is characterized by a mobility gap $\hbar \omega_c$. For
the Hall resistance of a system with $N$ completely filled LLs one
finds the value ${1 \over N} {h \over e^2}$, and as a function of the
chemical potential, the Hall resistance follows a stair step curve
with plateaus at exactly these values.

Impurities in real samples lead to a broadening of the delta function
peaks in the DoS, i.e., the disappearance of the excitation gap.  In
addition, electronic states become localized with exception of a
region of delocalized states centered around $E = E_n$.  As a
consequence, the excitation gap of the clean system is replaced by a
mobility gap in the disordered system, which leads to a stair step
curve of the Hall resistance as a function of the magnetic field or
the electron density.  Due to the topological nature of the quantum
Hall resistance as an integral over the Brillouin zone, the value of
the plateau resistance is unchanged as compared to the clean system
\cite{Thouless,physicstoday}.

Real samples are finite and have contacts. The sample boundaries are
described by a confining potential that prevents the electrons from
leaving the sample and leads to the formation of one--dimensional edge
states \cite{Halperin82}.  Theoretically, these edge states can be
modelled as ideal one--dimensional wires, and assuming the above
described localization--delocalization scenario for the bulk states,
one can derive the quantized Hall resistance in the framework of the
Landauer--B\"uttiker approach \cite{Buettiker88}.  The current
distribution in real experiments is not restricted to narrow channels
along the sample edges but generally involves delocalized bulk states
as well. For this reason, both the Hall resistivity of bulk states and
the resistance of edge states need to be combined for a complete
description of experimental results \cite{Hajdu}.

For reasons of simplicity, we will ignore both disorder and edge
effects first. We make a quantitative prediction of the influence of
gravitational corrections on the Hall resistance of a clean system
with a completely filled highest LL.  In addition, we present a
qualitative argument why the result of this calculation should be
valid in the presence of edge states and disorder as well.  The
influence of gravity up to order $g/c^2$ is described by the
Hamiltonian \cite{HehlNi}
%
%************************* gravitational field terms******************
\begin{equation}
  \hat{H}_{\rm grav} = - m \varphi(\hat{\bf x}) - {1 \over 2 m}\,
  \hat{\bf p}\, {\varphi(\hat{\bf x}) \over c^2}\cdot \hat{\bf p} -
  {\hbar \over 4 m c^2}\, \hat{\bf \sigma} \cdot ( {\bf \nabla}
  \varphi(\hat{\bf x}) \times \hat{\bf p}) \ \ .
\label{gravterms.eq}
\end{equation}
%********************************************************************
%
We first discuss the situation where ${\bf \nabla} \varphi({\bf x})$
is perpendicular to the plane of the 2DEG. Labeling the plane of the
2DEG as the $xy$--plane, the gravitational potential depends on $z$
only and hence commutes with $\hat{p}_x$ and $\hat{p}_y$.  The first
term in $\hat{H}_{\rm grav}$ turns into a constant which contributes
an oscillatory time dependence to the wave functions. The second term,
which describes the gravitational redshift of the kinetic energy, does
not depend on position any more and only modifies the effective mass
of the electrons, which drops out of the calculation of the Hall
resistance. The third term is analogous to the Rashba term
\cite{Rashba60} $\gamma \hat{\sigma} (-i \bf{\nabla} \times \bf{E})$,
which describes the influence of the confining electric field on the
electron spin. The Rashba term is not known to influence the accuracy
of the integer QHE. In addition, the coupling strength $\gamma E \sim
10^{-12} e V m$ is about 23 orders of magnitude larger than the
gravitational coupling $ {\hbar^2 g \over 4 m c^2} \sim 10^{-35} e V
m$.  In conclusion, there is no evidence that a gravitational field
perpendicular to the plane of the 2DEG influences the Hall resistance
to order $\varphi / c^2$.

Next, we assume ${\bf g}({\bf x})$ to be in the plane of the 2DEG.
The electric field needed to counteract the gravitational force is $E
= m g /e \sim 10^{-10}V/m$ and hence about ten orders orders of
magnitude smaller than the Hall electric field in metrological
applications. The influence of the gravitational field on the ions in
a conductor is stronger by a factor of $10^2$ and hence more important
for actual measurements. The ``gravitational Rashba term'' in
(\ref{gravterms.eq}) is again dominated by the corresponding term due
to the externally applied electric field.  To discuss the
gravitational redshift of the kinetic energy, we assume ${\bf g}$ to
be oriented along the negative $x$--direction and use a Taylor
expansion $\varphi({\bf x})= \varphi_0 - g x$.  We consider a torus
with finite extension $L_y$ in $y$--direction and infinite extension
in $x$--direction.

Using the Landau gauge ${\bf A}({\bf x}) = ( 0 , B x, 0)$, we make the
usual product ansatz of a plane wave of momentum $k$ in $y$--direction
and a $x$--dependent $\psi_\alpha(x)$, with $\alpha = (n, k)$ denoting
the set of quantum numbers.  Upon inserting this ansatz into the full
Schr\"odinger equation, the effective Hamiltonian for
$\psi_\alpha(x)$, in the presence of an electric field $E$ in
$x$--direction, reads
%
%************************* effective Hamiltonian *********
\begin{eqnarray}\hspace{-20pt}
  H_x & = & {1 \over 2} \hat{p}_x ({1 \over m} + {g x \over m c^2})
  \hat{p}_x + {1 \over 2} m \omega_c^2(x - X)^2 
\label{hameff.eq}\\ 
&+ &  {1 \over 2} {
    g m \omega_c^2 \over c^2} x (x - X - {e E \over m \omega_c^2})^2 +
  e E X + {e^2 E^2 \over 2 m \omega_c^2} \ . \nonumber
\end{eqnarray}
%*********************************************************
%
Here, $X = - {\hbar k \over e B} - {e E \over m \omega_c^2}$.  The
current in $y$--direction of a state $\Psi_\alpha$ can be calculated
as the derivative
%
%*********************** current equation ****************
\begin{equation}
  I_y = - e \langle \alpha | \hat{v}_y | \alpha \rangle = -(1/ e B)\,
  \partial \epsilon_\alpha / \partial X\ \ .
\end{equation}
%*********************************************************
%  
To order $O(g/c^2)$, a perturbative evaluation of the corresponding
terms in the Hamiltonian (\ref{hameff.eq}) is sufficient.  As we want
to discuss the influence of disorder and a boundary potential, we
first use a transformation which turns the position dependent
effective mass (first term in (\ref{hameff.eq}))
%
%************* position dependent effective mass**********
\begin{equation}
  {1/ m(x)} = {1/ m} + {g x/( m c^2)} \ \ 
\end{equation}
%*********************************************************
%
into an effective potential.  Following Gonul et al.\cite{Go+02}, the
position dependent effective mass is replaced by a constant mass $m_0$
by applying a coordinate transformation $x = f(\tilde{x})$ whose
inverse is defined by $\tilde{x} = \int_0^x du\sqrt{m(u)/m_0} =
f^{-1}(x)$.  This coordinate transformation leaves the spectrum of the
Hamiltonian unchanged but renormalizes the wave function and changes
the potential $V(x)$ to an effective potential
$V_{eff}(\tilde{x})$. To order $O(g/c^2)$, we find
%
%********************** change in potential *************
\begin{equation}
  V_{eff}(\tilde{x}) = V(f(\tilde{x})) + O((g/c^2)^2) \ \ .
\label{effpot.eq}
\end{equation}
%********************************************************
%
An evaluation of the energy shift due to the gravitational corrections
yields the  current density
%
%****************************current formula ************
\begin{equation}
  j_y = {e^2 \over h} E_x \left( 1 + {g \over c^2} {e E_x \over m
      \omega_c^2} \right) + {e g \over h c^2} \hbar \omega_c
  (n+{1\over 2}) 
\end{equation}
%********************************************************
%
for a completely filled $n$--th LL.  The constant background current
(last term) is, with $10^{-23} A$, much smaller than the typical
experimental currents of about $10^{-7} A$.  The nonlinear correction
to the Hall conductivity for typical values of the electric field is
about $10^{-19}$ and does not influence experimental results.

The change of the disorder potential due to the transformation
(\ref{effpot.eq}) is reflected in the change of the correlation
function for the disorder potential. However, as the details of the
disorder correlator are known to be irrelevant for the derivation of
quantum Hall plateaus, we conjecture that the gravitational terms do
not change the localization--delocalization scenario responsible for
the integer QHE.  Similarly, the formation of edge states does not
depend on the details of the confining potential, and we find it
unlikely that the qualitative properties of edge states are changed by
the transformation (\ref{effpot.eq}).

In summary, our calculations suggest that the linear Hall resistance
is not influenced by a gravitational field to order $O(g/c^2)$. This
finding corroborates the idea arising from a macroscopic argument that
the Hall resistance may be completely independent of the gravitational
field.  For a field orientation parallel to the 2DEG, we find both a
constant background current and a nonlinear contribution to the Hall
current.  The only term which is possibly relevant for experiment is
the background voltage caused by the gravitational potential.  This
contribution could be detected with an experimental accuracy of
$10^{-8}$.

{\it Acknowledgments.} We appreciate a discussion that we had with K.\
von Klitzing (Stuttgart) on the QHE and, in particular, on its
possible dependence on the gravitational field. Moreover, we thank
R.~Klesse (Cologne) for valuable remarks and M.~Kelly (NIST) for
providing helpful information.  This project has been supported by the
grant HE 528/20-1 of the DFG (Bonn).

\end{document}